\documentclass[useAMS,usenatbib]{mn2e}
\newcommand{\Tcnmax}{\mathfrak{T}_\mathrm{cn}^\infty}

\usepackage{amssymb}
\usepackage{epsf}
\usepackage{bm}

\title[Superfluid modes in oscillating neutron stars]
{Nonradial superfluid modes in oscillating neutron stars}
\author[A. I. Chugunov, M. E. Gusakov]
{A.~I.~Chugunov\thanks{andr.astro@mail.ioffe.ru},
M. E. Gusakov\thanks{gusakov@astro.ioffe.ru}\\
Ioffe Physical-Technical Institute of the Russian Academy of
Sciences, Politekhnicheskaya 26, 194021 Saint-Petersburg, Russia}

\begin{document}

\date{Accepted 2011 xxxx. Received 2011 xxxx;
in original form 2011 xxxx}

\pagerange{\pageref{firstpage}--\pageref{lastpage}} \pubyear{2011}

\maketitle

\label{firstpage}

%
\begin{abstract}
For the first time nonradial oscillations of superfluid nonrotating
stars are self-consistently studied at finite stellar temperatures.
We apply a realistic equation of state and realistic density dependent model of
critical temperature of neutron and proton superfluidity.
In particular, we discuss three-layer configurations of a star with no
neutron superfluidity at the centre and in the outer region of the
core but with superfluid intermediate region.
We show, that oscillation spectra contain
a set of modes whose frequencies can be
very sensitive to temperature variations.
Fast temporal evolution of the pulsation spectrum in the course of
neutron star cooling is also analysed.
\end{abstract}
%


\maketitle

\section{Introduction}

Studying the pulsations of neutron stars (NSs)
is very important and actively developing area of research,
since comparison of pulsation theory with observations
can potentially give valuable information about the properties
of superdense matter (\citealt{afjkkrrz11}).
Yet, it is an extremely difficult theoretical problem
even for normal (nonsuperfluid) stars.
Superfluidity of baryons additionally complicates the theory
because in superfluid (SFL) matter one should deal with
several independent velocity fields.
As a consequence,
the hydrodynamics describing
pulsations of SFL NSs is much
more complicated in comparison
to ordinary (normal) hydrodynamics (e.g., \citealt{BulkVisc}).

In this letter we report on a substantial progress in
modeling and understanding the nonradial oscillations
of nonrotating SFL NSs
in full general relativity.
It should be noted that
the nonradial oscillations of such stars
have been intensively studied in the literature
starting from the seminal paper by \cite{LM94}.
In particular, nonradial oscillations of general relativistic NSs
were considered by \cite*{cll99, acl02, yl03, SF_oscil}.
Because of the complexity of the problem,
most of these papers used
a simplified microphysics input, i.e.,
toy-model equations of state
and simplified models of baryon superfluidity
(see, however, \citealt{SF_oscil}; \citealt*{hap09}; \citealt{ha10}).
Moreover, in {\it all} these studies
SFL matter was treated
as being at zero temperature,
an assumption that is unjustified
at not too low stellar temperatures
and may lead to a quantitatively incorrect oscillation spectra
(\citealt{GA, RadPuls}; this letter).

Here we improve on this
by considering
nonradial oscillations of SFL neutron stars
at {\it finite} temperatures.
We follow an approach of \cite{GK11} (hereafter GK11)
which allows us to analyse
oscillations of general relativistic NSs
employing realistic equation of state,
density dependent profiles of
nucleon critical temperatures and fully
relativistic finite-temperature SFL hydrodynamics.

As it was first found by \cite{LM94}, oscillation spectrum of a SFL
NS consists of two distinct classes of modes, the so called normal
and superfluid modes. The frequencies of normal modes almost
coincide with the oscillation frequencies of a normal star and hence
are independent of temperature. The spectrum of these modes is
therefore very well studied in the literature (see, e.g.,
\citealt*{ThornPaperI, Osc1, bfg04}). On the contrary, SFL modes can
be very temperature-dependent. Using the approach of GK11 they can
be decoupled from the normal modes and studied separately. Since
radial SFL modes have already been thoroughly analysed in
\cite{RadPuls}, here we focus on the nonradial SFL modes. In what
follows, the speed of light $c=1$.

\section{Basic equations} \label{Sec:BasicEq}

In this section we briefly
discuss the equation describing SFL oscillation modes
[Eq.\ (\ref{main})].
The detailed derivation of this equation
can be found in GK11.
We consider a NS with nucleonic core
and assume that both neutrons (n) and protons (p)
can be superfluid.
Following GK11, we introduce
the baryon current density
$j_{(\mathrm b)}^\mu=j_{(\mathrm n)}^\mu+j_{(\mathrm p)}^\mu$,
where
\begin{equation}
j_{(i)}^\mu=n_i\, u^\mu+Y_{ik} w_{(k)}^\mu
\label{jjj}
\end{equation}
is the current density for particles $i={\rm n}$ or ${\rm p}$
(e.g., \citealt{RadPuls}). 
Here and below the summation is assumed over the
repeated nucleon index $k=\mathrm n$, $\mathrm p$. In Eq.\
(\ref{jjj}) $n_i$ is the number density; $u^\mu$ is the
four-velocity of ``normal''\ liquid component; $w_{(k)}^\mu$ is the
four-vector that characterizes motion of superfluid neutron
($k=\mathrm n$) or proton ($k=\mathrm p$) component with respect to
normal matter. Finally, the symmetric temperature-dependent matrix $Y_{ik}$ ($=Y_{ki}$) 
is a relativistic analogue of the SFL entrainment matrix. 
Since electrons $(\mathrm e)$ are normal, their current
density is $j_{(\mathrm e)}^\mu=n_\mathrm e\, u^\mu$, where
$n_\mathrm e$ is the electron number density. The quasineutrality
implies $n_\mathrm e=n_\mathrm p$ (for simplicity, we ignore
possible admixture of muons in the core).

In this letter we study small-amplitude (linear) oscillations of
a {\it nonrotating} star being initially in hydrostatic equilibrium. 
Hence, for the unperturbed star one has 
$u^{\mu}=({\rm e^{-\nu/2}},0,0,0)$ 
and $w_{(\mathrm n)}^\mu=w_{(\mathrm p)}^\mu=0$,
while the metric is 
    $\mathrm d S^2=-\mathrm e^\nu \mathrm d t^2
    +\mathrm e^\lambda \mathrm d r^2+r^2 \mathrm d \Omega^2$,
where $r$ and $t$ are the radial and time coordinates, respectively;
$\nu(r)$ and $\lambda(r)$ are the metric functions; 
and $\Omega$ is the solid angle
in a spherical frame with the origin at the stellar centre.
For definiteness,
we assume that the unperturbed matter in the stellar core
was in beta-equilibrium, 
$\delta \mu \equiv \mu_\mathrm n-\mu_\mathrm p-\mu_\mathrm e=0$, 
where $\mu_d$ is the chemical potential for particles $d=\mathrm{n}$, p, e. 
We also restrict ourselves to oscillations with vanishing
electrical current, $j_{(\mathrm p)}^{\mu}-j_{(\mathrm e)}^{\mu}=0$.
The latter condition couples the SFL degrees of freedom,
\begin{equation}
 w_{(\mathrm p)}^\mu=-(Y_{\mathrm{pn}}/Y_{\mathrm{pp}})\,w_{(\mathrm n)}^\mu.
\label{NoCurrent}
\end{equation}

As it was shown in GK11
the interaction between the SFL and
normal oscillation modes is controlled by the coupling parameter $s$,
which is given by 
$s=(n_{\rm e} \, \partial P/\partial n_{\rm e})
/(n_{\rm b} \,\partial P/\partial n_{\rm b})$
where $P(n_{\rm b}, n_{\rm e})$ is the pressure 
and $n_{\rm b}=n_{\rm n}+n_{\rm p}$
is the baryon number density.
This parameter is small for a wide set of realistic equations of state,
$\left|s\right|\lesssim 0.05$
(see fig.\ 1 in GK11),
so that the approximation of completely decoupled SFL and
normal modes ($s=0$) is already sufficient to calculate the pulsation spectrum
within an accuracy of a few per cent.
In the $s=0$ approximation
the quantities $j^{\mu}_{(\mathrm b)}$, $P$, 
and the metric $g_{\mu \nu}$
remain unperturbed for
SFL oscillation modes.
This opens up a possibility to formulate an equation
describing SFL modes
that depends {\it only} on SFL degrees of freedom,
i.e., on $w^{\mu}_{(k)}$.
This equation follows from the energy-momentum
conservation and potentiality condition for motion of SFL neutrons.
For a nonrotating star it takes the form (see GK11)
\begin{equation}
    \imath \omega\left( \mu_n Y_{\mathrm n k} w_{(k)j}-n_\mathrm b
    w_{(\mathrm n) j}\right)=n_\mathrm e \,
    \partial_j \left(\delta \mu^\infty \right), \label{sf_eq}
\end{equation}
where $\delta \mu^\infty \equiv \delta \mu \, {\mathrm e}^{\nu/2}$
and the disbalance $\delta \mu(n_{\mathrm b}, n_{\mathrm e})$
equals
\begin{equation}
     \delta \mu
     =-\imath\, \mathrm e^{\nu/2}\,\frac{\mathfrak{B} \, n_\mathrm e}{\omega}
     \left[\frac{Y_{\mathrm n k}}{n_\mathrm e\, n_\mathrm b}
     \frac{\partial n_\mathrm e}{\partial x^\mu}\, w_{(k)}^\mu +\left(\frac{Y_{\mathrm n k}}{n_\mathrm b}
     w_{(k)}^\mu\right)_{;\mu}\right]. \label{dmu}
\end{equation}
In Eqs.\ (\ref{sf_eq}) and (\ref{dmu})
$j=1$, 2, or 3 is the space index,
$\partial_j \equiv \partial/(\partial x^j)$,
$\mathfrak{B} \equiv \partial \delta \mu(n_{\mathrm b}, n_{\mathrm e})/\partial n_\mathrm e$,
and all perturbations are assumed to be $\propto \exp(\imath\omega t)$.
Because Eq.\ (\ref{sf_eq}) is linear,
one can generally present
$\delta \mu^\infty$ as:
$\delta \mu^\infty(r,\Omega)=\delta \mu_l(r)\,{\rm Y}_{lm}(\Omega)$,
where ${\rm Y}_{lm}(\Omega)$ is the spherical harmonic
[notice that $\delta \mu_l(r)$ does not depend on the index $m$, see Eq.\ (\ref{main})].
Combining then
Eqs.\ (\ref{NoCurrent}), (\ref{sf_eq}), and (\ref{dmu})
one
obtains the following equation
\begin{eqnarray}
    0 &=& \delta
    \mu_l^{\prime\prime}+\left(\frac{h^\prime}{h}
-\frac{\lambda^\prime}{2}
    +\frac{2}{r}\right)\delta
    \mu_l^\prime 
    \nonumber\\
    &-&\mathrm e^\lambda\left[\frac{l(l+1)}{r^2}+\mathrm
    e^{-\nu/2}\frac{\omega^2}{h \, \mathfrak{B}}
    \right] \delta \mu_l. 
    \label{main}
\end{eqnarray}
%
%
%
Here prime means derivative with respect to $r$ and
$h=e^{\nu/2} \,n_{\rm e}^2/(\mu_{\rm n} \,n_{\rm b} \,y)$, 
where 
$y=n_{\rm b} \, Y_{\rm pp}/[\mu_{\rm n}\, (Y_{\rm nn} Y_{\rm pp}-Y_{\rm np}^2)]-1$.

Eq.\ (\ref{main})
determines the eigenfrequencies of SFL oscillation modes
and is valid in the region of the stellar core where neutrons are superfluid
(hereafter, SFL-region).
This equation depends on the internal stellar temperature $T$
only through the parameter $y=y(T/T_{\rm cn},T/T_{\rm cp})$,
where $T_{{\rm c}i}(r)$ is the profile of critical temperatures for
particles $i={\rm n}$ or p. 
The boundary conditions to Eq.\ (\ref{main}) are following.
If neutrons in the stellar centre are superfluid,
the regularity of the
solution requires $\delta \mu_l\propto r^l$ at $r\rightarrow 0$.
If the outer boundary of the SFL-region
coincides with the crust-core interface
(where $r=R_{\rm cc}$),
then
the
SFL current should not penetrate the crust.
This condition implies
$\delta \mu_l^\prime(R_{\rm cc})=0$.
Finally, if 
$T$ is so high,
that the SFL-region does not spread all over the core,
then
its boundaries
are determined by the condition $T=T_\mathrm{cn}(r)$
(see Sec.\ \ref{Sec:micro} for more details).
In that case, the regularity of the solution at such boundaries requires
$\delta \mu_l^\prime=\mathrm e^{\lambda-\nu/2}\, \omega^2 \, 
\delta \mu_l/(h^\prime \, \mathfrak{B})$.

Thus, we reduce the problem of calculation of SFL modes
to solving simple
second-order differential Eq.\ (\ref{main}).
For any fixed multipolarity $l$ the solution to 
Eq.\ (\ref{main})
consists of a set of
eigenfrequencies $\omega_{ln}$
and eigenfunctions $\delta \mu_{ln}(r)$
which differ
by the number
of radial nodes $n=0$, 1, 2, $\ldots$

\section{Microphysics input and NS model}
\label{Sec:micro}
Prior to studying the SFL oscillations using Eq.\ (\ref{main}), one
has to specify equation of state 
(including profiles of $T_{\rm cn}$ and $T_{\rm cp}$) 
and construct a hydrostatic model of an
unperturbed star.
In addition, one has to specify the
profile of internal stellar temperature.
High thermal conductivity leads to a rapid equilibration of $T$ in
the NS core
(see, e.g., \citealt*{gyp01}).
As a result, the red-shifted internal temperature
$T^\infty=T\mathrm e^{\nu/2}$
becomes almost constant.
Moreover, as it was shown by \cite{GA},
for the SFL-region to be in hydrostatic and beta-equilibrium
it must be in thermal equilibrium.
Thus, in what follows we assume that $T^\infty={\rm const}$
in the SFL-region.

\begin{figure}
    \begin{center}
        \leavevmode
        \epsfxsize=3.2in \epsfbox{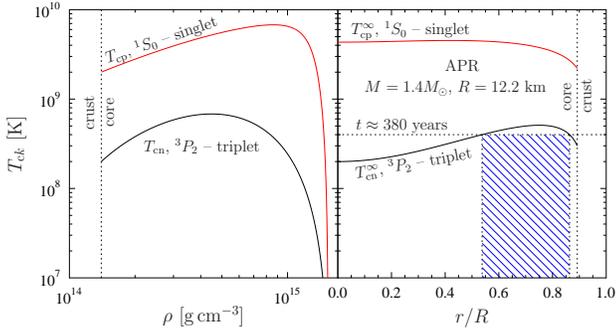}
    \end{center}
    \caption{(color online)
    Left panel: Nucleon critical temperatures $T_{\mathrm c k}$
    versus density $\rho$ ($k={\rm n}$, ${\rm p}$).
    Right panel: Red-shifted critical temperatures
    $T^\infty_{\mathrm c k}$ versus radial coordinate $r$.
    }
    \label{Fig:Tcr}
\end{figure}

In the present letter we employ the equation of state
suggested by \cite*{APR} (APR).
The coupling parameter $s$ for such an equation of state
is small, $|s| \sim 0.02$.
The density profiles of $T_{\rm cn}$ and $T_{\rm cp}$ 
that we use here are shown in the left panel of Fig.\ \ref{Fig:Tcr}.
They do not contradict to results of microscopic calculations (see, e.g., \citealt{ls01})
and are similar to the nucleon pairing models used 
to explain observations of the cooling NS 
in Cas A supernova remnant (\citealt{Sht11}).
For definiteness, all calculations are performed for
a star of the mass $1.4 M_\odot$
and circumferential radius $R=12.2$~km.

Since $T^\infty$, but not $T$, is constant in the SFL-region,
it is convenient to introduce
the {\it red-shifted} nucleon critical temperatures
$T^\infty_{\mathrm ck} \equiv {\rm e}^{\nu/2}\, T_{\mathrm ck}$
($k={\rm n}$, ${\rm p}$)
to analyse how the size of SFL-region changes with $T^\infty$.
The functions $T^\infty_\mathrm{cn}(r)$ and
$T^\infty_\mathrm{cp}(r)$ are shown on the right panel
of Fig.\ \ref{Fig:Tcr}.
The red-shifted proton critical temperature is high,
$T^\infty_{\rm cp}(r) \sim 2 \times 10^9$~K,
so that superfluid protons occupy the
entire core almost immediately after the NS birth.
The function $T^\infty_\mathrm{cn}(r)$ has a maximum 
$\Tcnmax\approx 5.1 \times 10^8$~K 
at $r=r_\mathrm{cn}^\mathrm{max}\approx 0.75 R$.
Near the stellar centre the density varies
slowly with $r$ which results in a
weak dependence of $T^\infty_\mathrm{cn}$
on the radial coordinate.
As the star cools down to $T^\infty \lesssim \Tcnmax$,
the SFL-region is
formed, initially, as a narrow spherical layer.
Upon subsequent cooling the layer becomes
wider and, for example, 
at $T^\infty=4\times10^8$~K
it is shown by
the hatched region in the figure.
As the temperature decreases further,
the SFL-region extends to the crust
and, eventually, 
at $T^\infty=T_\mathrm{cn}^\infty(0)\approx 2 \times 10^8$~K 
it penetrates the stellar centre.

\section{Oscillation spectra and modes}

\begin{figure}
    \begin{center}
        \leavevmode
        \epsfxsize=2.9in \epsfbox{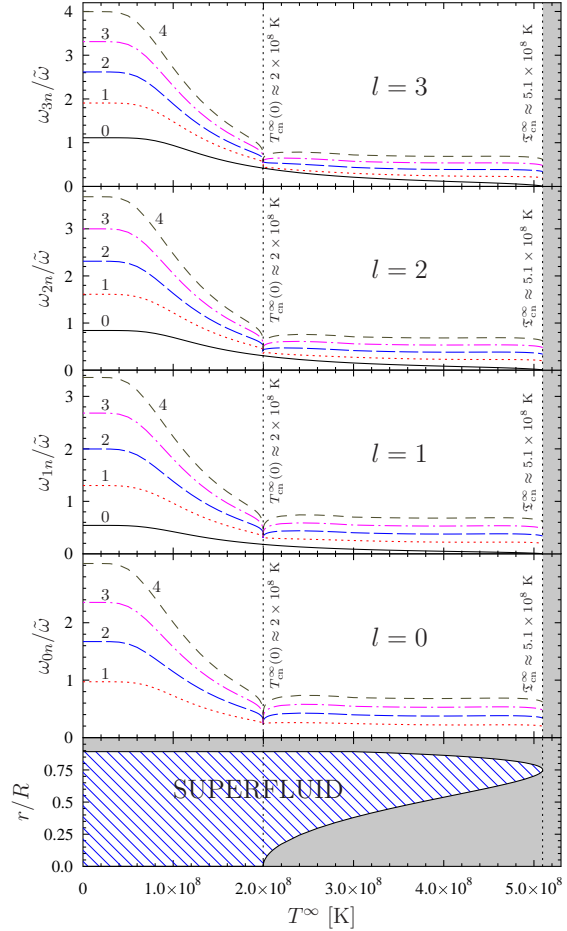}
    \end{center}
    \caption{(color online) Eigenfrequencies $\omega_{ln}$ versus $T^\infty$
    for multipolarities $l=0$, $1$, $2$, and $3$.
    For each $l$ a set of curves is plotted with $n=0$, $1$, $2$, $3$ or $4$.
    At $T^\infty \leq T_{\rm cn}(0) \approx 2 \times 10^8$~K
    (see the vertical short-dashed line)
    superfluidity occupies the stellar centre.
    The bottom panel shows variation of SFL-region with $T^{\infty}$.
    For more details see the text.
    }
    \label{Fig:Spec}
\end{figure}
%
Figure \ref{Fig:Spec} 
presents normalized eigenfrequencies $\omega_{ln}$
(in units of $\tilde{\omega}=c/R\approx2.5\times10^4$~s$^{-1}$)
versus internal temperature $T^\infty$ for SFL oscillations
of multipolarity $l=0$, 1, 2, and 3.
For each $l$ we plot a set of oscillation modes
that differ by the number of radial nodes $n=0$ (solid lines), $n=1$ (dots),
$n=2$ (long dashes), $n=3$ (long-short dashes), and $n=4$ (dashes).
One sees that the higher the $n$ the larger the $\omega_{ln}$.
By the hatches in the
bottom panel of the figure
we show the SFL-region;
as expected, the size of this
region depends on $T^\infty$.
The grey-shaded area corresponds to
the crust and a region in the core
where all neutrons are unpaired.
A similar shaded area on the four upper panels shows temperatures
$T^\infty\ge \Tcnmax$
for which {\it all} neutron matter in the core is normal.
In the latter case there are no
SFL modes in NS.

\begin{figure*}
    \begin{center}
        \leavevmode
               \epsfxsize=6.82in \epsfbox{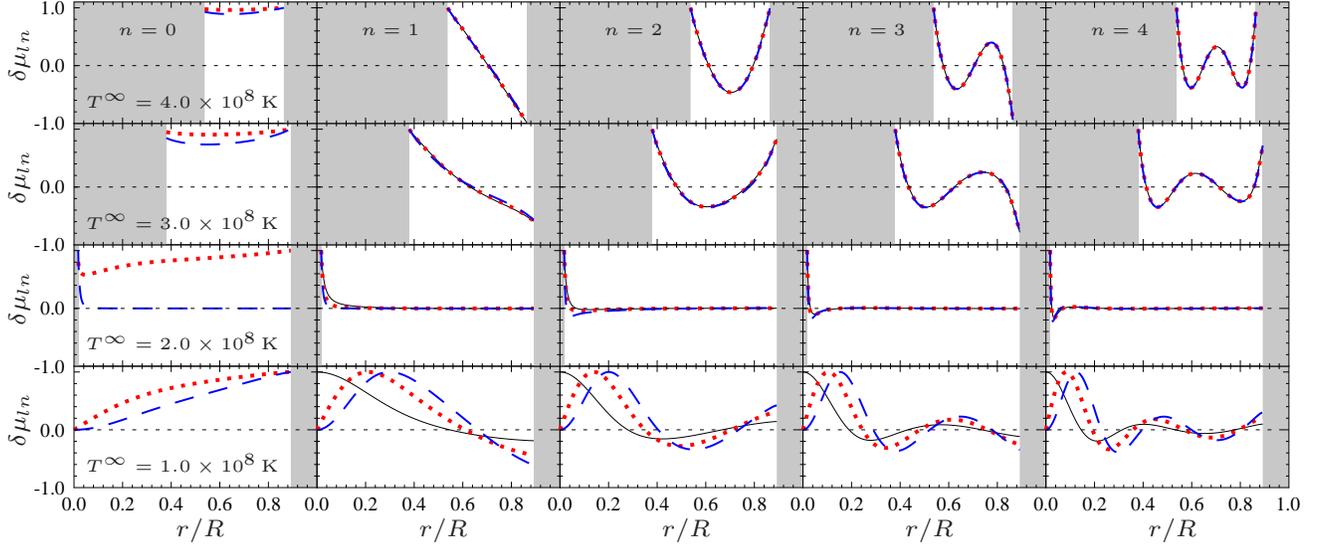}
    \end{center}
    \caption{(color online) Eigenfunctions $\delta \mu_{ln}$ versus $r$ for
    $T^\infty=10^8$, $2 \times 10^8$, $3 \times 10^8$, and $4 \times 10^8$~K
    (5 panels in a row for each temperature).
    The columns of panels are for $n=0,$ $1$, $2$, $3$, and $4$ radial nodes.
    Solid, dotted, and dashed lines correspond to multipolarities
    $l=0$, $1$, and $2$, respectively. For more details see the text.
    }
    \label{Fig:eigen}
\end{figure*}

Before further discussing spectra in Fig.\ 2
it is convenient to describe briefly Fig.\ 3 that
presents eigenfunctions $\delta \mu_{ln}(r)$
normalized to unity in the maximum.
The solid lines correspond to
radial oscillation modes ($l=0$), dotted and dashed lines describe
dipole ($l=1$) and quadrupole ($l=2$) modes, respectively.
Each column
in the figure contains four panels which are plotted for the
following temperatures (from bottom to the top): 
$T^\infty=10^8$, $2 \times 10^8$, $3\times 10^8$, and $4\times 10^8$~K.
For
any of these temperatures
we have five panels in a row, which correspond to
(from left to right) $n=0$, $1$, $2$, $3$,
and $4$ radial nodes of $\delta \mu_{ln}(r)$.
The stellar regions where neutrons are normal, are shaded in
Fig.\ \ref{Fig:eigen}.

\begin{figure*}
    \begin{center}
        \leavevmode
        \epsfxsize=6.82in \epsfbox{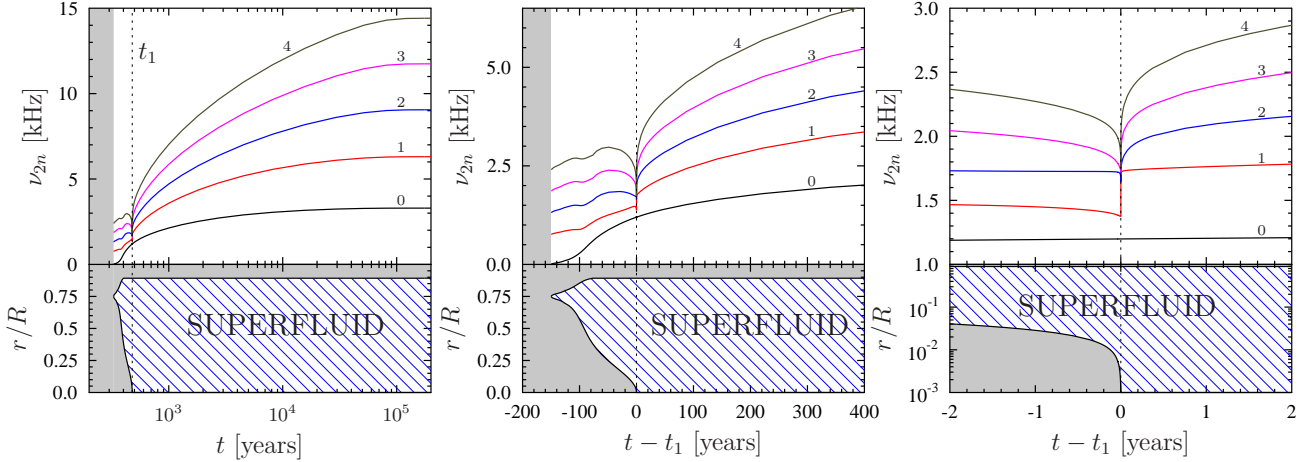}  
    \end{center}
    \caption{(color online) Upper panels:
    The oscillation frequencies $\nu_{2n}$ of quadrupole modes with $n=0$,$\ldots$,$4$
    versus stellar age $t$.
    Vertical dashed lines show the stellar age $t_1$ at which neutron superfluidity
    first appears in the stellar centre.
    Bottom panels: Variation of SFL-region with $t$.
    For more details see the text and captions
    to Figs.\ \ref{Fig:Spec} and \ref{Fig:eigen}.}
    \label{Fig:l2evol}
\end{figure*}

At
$T^\infty\lesssim \Tcnmax$
the size of SFL-region rapidly increases as
the star cools down,
whereas the eigenfrequencies
are almost temperature-independent for $T^\infty \gtrsim 3 \times 10^8$~K
(except for the modes with $n=0$, see Fig.\ 2 and discussion below).
This is so
because of compensation of two opposite tendencies:
(i) expansion of the SFL-region
and
(ii) increasing of the local speed of SFL sound
$v^\mathrm{sf} = {\rm e}^{-\nu/4} \sqrt{-h\,\, \mathfrak{B}}$
with decreasing $T^\infty$
(see \citealt{GA} for details on $v^\mathrm{sf}$).
The tendency (i) leads to decreasing
while (ii) leads to growing of the eigenfrequencies $\omega_{ln}$.

The qualitative behaviour of $\omega_{ln}$ 
and $\delta \mu_{ln}(r)$ at $3\times10^8$~K$\lesssim
T^\infty\lesssim \Tcnmax$
follows from simple arguments. 
The parameter $h$ is small there,
so that it is sufficient to retain
only the terms $\propto h^{-1}$
in Eq.\ (\ref{main}).
As a result,
for $T^\infty\gtrsim 3\times10^8$~K
$\delta \mu_{ln}(r)$
will be almost independent of $l$
(see Fig.\ \ref{Fig:eigen}).
Furthermore, at $T^\infty\gtrsim 4\times10^8$~K
the function $h$ can be approximated as
$h\approx A\left[(\Tcnmax-T^\infty)
-B(r-r_\mathrm{cn}^\mathrm{max})^2\right]$,
where $A$ and $B$ are some constants
depending on a NS model.
Using this approximation,
one can analytically solve Eq.\ (\ref{main})
and find that eigenfunctions
$\delta \mu_{ln}(r)$ are proportional
to the Legendre polynomials $P_n$ while
the eigenfrequencies $\omega_{ln} \propto \sqrt{n(n+1)}$
and are independent of $T^\infty$
(see Figs.\ \ref{Fig:Spec} and
\ref{Fig:eigen} for $T^\infty\gtrsim 4\times10^8$~K).
For the modes with $n=0$
this estimate
gives $\omega_{l0}=0$.
As follows from Fig.\ \ref{Fig:Spec},
in that case the eigenfrequencies are indeed small but nonzero.
To improve the estimate for $\omega_{l0}$
one should take into account
the term depending on $l(l+1)/r^2$ in Eq.\ (\ref{main}).
Because at $T^\infty \gtrsim 4 \times 10^8$~K 
$\delta \mu_{l0}(r)$
is almost constant (see top-left panel in Fig.\ \ref{Fig:eigen}), 
one can put $\delta \mu_{l0}'\approx 0$ and $\delta \mu_{l0}'' \approx 0$.
Using then Eq.\ (\ref{main}),
one obtains the following estimate for $\omega_{l0}$:
$\omega_{l0} \propto \sqrt{l\,(l+1)}\,v^\mathrm{sf}/r_\mathrm{cn}^\mathrm{max}\propto
\sqrt{l\,(l+1)}\,\sqrt{1-T^\infty/\Tcnmax}$.
This formula indicates that $\omega_{l0}$
depends only on $T^\infty/\Tcnmax$
(but not on the size of SFL-region)
and vanishes at
$T^\infty = \Tcnmax$ (see Fig.\ \ref{Fig:Spec}).
It remains to note that for $l=0$
Eq.\ (\ref{main}) has a static solution
$\delta \mu_{00}= \delta \mu^\infty=\mathrm{const}$ and $\omega_{00}=0$,
which describes a star in hydrostatic and diffusive equilibrium 
(but not necessarily in beta-equilibrium, see \citealt{GA} for more details).
It should also be stressed that
approximate formulas for eigenfrequencies found above
is not a special feature of our microphysical model.
Rather, it is
inherent in {\it all} NSs for which the maximum of
$T^\infty_\mathrm{cn}(r)$ lies between the centre and the crust-core
boundary.

At $T^{\infty}$ slightly exceeding $T_{\rm cn}^\infty(0)$
the size of SFL-region rapidly expands to the stellar centre
as the star cools down.
The reason for that is weak dependence of
$T^\infty_\mathrm{cn}$ on $r$ at $r\lesssim 0.1\, R$
(see the right panel of Fig.\ \ref{Fig:Tcr}).
Since $T^\infty \approx T^\infty_\mathrm{cn}$
for $r\lesssim 0.1\, R$,
$v^\mathrm{sf}$ is close to 0 there.
This results in a noticeable decrease
of $\omega_{ln}$ with $n\geq 1$ near $T_{\rm cn}^\infty(0)$
(see Fig.\ \ref{Fig:Spec}). 
In particular, at $T^\infty=2 \times 10^8$~K 
the corresponding oscillation modes
are all localized in the vicinity of the stellar centre,
where $v^\mathrm{sf}$ is small (see Fig.\ \ref{Fig:eigen}).
During subsequent cooling
[for $T^\infty\lesssim T_\mathrm{cn}^\infty(0)$]
the entire core is already occupied by the
neutron superfluidity and
increasing of $v^\mathrm{sf}$,
especially in the central regions of the star,
leads to growing of $\omega_{ln}$.
When $T^{\infty}$ drops below $\sim 4\times 10^7$~K the
eigenfrequencies $\omega_{ln}$ and eigenfunctions $\delta \mu_{ln}(r)$ become almost
independent of $T^\infty$,
since $h$ (and $v^\mathrm{sf}$)
approach their asymptotes at $T^\infty=0$.

Let us now discuss the temporal evolution
of oscillation spectra during the star cooling.
To simulate cooling we apply a slightly updated version
of the code discussed by \cite{Cool1}.
The onset of neutron superfluidity is accompanied
by an abrupt acceleration of cooling
due to the Cooper pairing neutrino emission process
(see \citealt*{Sht11} for recent observational
evidences of such cooling).
As is demonstrated in Fig.\ \ref{Fig:l2evol},
this process leads to a rapid evolution of oscillation spectrum.
Upper panels in Fig.\ \ref{Fig:l2evol}
show the eigenfrequencies $\nu_{2n} \equiv \omega_{2n}/(2 \pi)$
of quadrupole modes ($n=0,\ldots,4$)
versus NS age $t$.
The lower panels illustrate change of the
SFL-region with $t$.
On the left upper panel the functions $\nu_{2n}(t)$
are plotted for the time interval
from $200$ to $2\times 10^5$ years.
After the onset of neutron superfluidity
(at $t \approx 330$ years)
the oscillation spectrum rapidly changes and
approaches its zero-temperature
asymptote only at $t\gtrsim 10^5$ years.
At the star age $t_1\approx 480$~years
superfluidity penetrate the stellar centre
[this corresponds to $T^\infty=T_{\rm cn}^\infty(0)$].
The period of time $|t-t_1| \ll t_1$
is accompanied by a very fast
evolution of the spectrum.
In more detail the evolution is shown in the central panels which
span an interval of approximately 600 years of NS cooling.
The time $t$ in these panels is counted from $t_1$.
The right panels in Fig.\ \ref{Fig:l2evol} show the only four-year
episode of NS life in the vicinity
of $t_1$.
One sees that even for such a small
period of time
the frequencies of some oscillation modes increase by more than
10\%, whereas the frequency of the mode with four radial nodes
changes noticeably on a time-scale of a few months~(!).

\section{Summary and outlook}

In this letter we study, for the first time,
nonradial oscillations of SFL NSs at finite temperatures.
Use of an approach developed in GK11
enable us to solve this problem in full general relativity,
employing the realistic equation of state (APR) and realistic,
density-dependent profiles of nucleon critical temperatures (Fig.\ \ref{Fig:Tcr}).
It is shown that
equations describing
SFL modes
can generally be reduced to
the simple second-order differential Eq.\ (\ref{main}).
The eigenfrequency spectrum (Fig.\ \ref{Fig:Spec})
and eigenfunctions (Fig.\ \ref{Fig:eigen}) for this
equation are carefully analysed.
It is demonstrated that dependence of eigenfrequencies $\omega_{ln}$
on $T^\infty$
is determined by two competing effects:
(i) decreasing of $\omega_{ln}$ with expanding SFL-region and
(ii) growing of $\omega_{ln}$
with increasing of the local speed of SFL sound $v^{\rm sf}$.
These results agree with the conclusions made
by Kantor and Gusakov (2011)
for a radially oscillating 
NS.

In addition, we examine the evolution
of oscillation spectrum in the course of NS cooling
(Fig.\ \ref{Fig:l2evol}).
We find that acceleration of NS cooling
soon after the onset of triplet neutron superfluidity
leads to a very fast modification of the spectrum
-- the eigenfrequencies can vary dramatically on a
time-scale of months.

In the present work we completely 
ignore various dissipative effects in pulsating SFL NSs.
Meanwhile, our results provide a strong basis to study 
these effects 
using the 
perturbative scheme 
suggested in GK11.
In particular, to determine the damping time $\tau$
of some normal or SFL mode
one could proceed in the following steps:
(i) find the vectors $u^{\mu}$ and $w^{\mu}_{(\rm n)}$ 
assuming $s=0$ and neglecting dissipation (see GK11 and Sec.\ 2); 
(ii) use them to calculate the dissipative terms entering the SFL hydrodynamics 
(and hence the rate of change $\dot{E}$ of the oscillation energy $E$, see, e.g., \citealt{RadPuls});
(iii) calculate $\tau$ as $\tau=-E/(2 \dot{E})$.

The other important problem concerns 
gravitational radiation from pulsating SFL NSs.
The radiation from normal modes
can be accurately calculated already in the $s=0$ limit
(and will be the same as that for a nonsuperfluid NS).
In turn, SFL modes are not coupled with the metric
in the $s=0$ limit,
so that to calculate gravitational radiation from them 
one must use the first-order perturbation theory in $s$.
Interestingly, this problem
is equivalent to finding gravitational
radiation from a {\it normal} NS experiencing
oscillations under an action of a small external force $\propto s$.
An intensity of such radiation will be reduced by a factor of
$s^2\sim 10^{-3}$ in comparison to normal modes
with the same $E$.
Detailed studies of
dissipation in SFL NSs are a
very important task, 
e.g., for calculation of the instability windows of r-modes;
we plan to address it in the near future.

\section*{Acknowledgments}
The authors are grateful to E.M.~Kantor, D.G.~Yakovlev, 
and A.V.~Brillante for useful comments. 
This work was partially supported by RFBR (grant 11-02-00253-a), 
by RF president program (grants NSh-3769.2010.2 and MK-5857.2010.2), 
and by the Dynasty foundation.


\label{lastpage}
\end{document}